%% file: article_20Ne_SR_rev.tex
\documentclass[aps,prc,twocolumn,showpacs,superscriptaddress,groupaddress,amsmath,amssymb]{revtex4-1}
\usepackage{graphicx}  
\usepackage{dcolumn}   
\usepackage{bm}        
\usepackage{verbatim}   
\usepackage{textcomp}

\usepackage{color}

\begin{document}

\title{Investigation of Direct Nuclear Reactions in a Storage Ring Using In-Ring Detection}


\input author_list.tex       

\date{\today}

\begin{abstract}
\begin{description}
\item[Background] Experiments involving nuclear reactions in a storage ring offer exceptional possibilities for precise measurements in inverse kinematics. These experiments provide excellent angular and energy resolution by particle spectroscopy, in addition to high luminosities. However, the extremely low-pressure environment maintained in the storage rings poses significant difficulties for experiments employing detectors or any outgassing material in the ring.

\item[Purpose] To investigate  nuclear reactions in inverse kinematics using the storage-ring technique. The reactions were induced by scattering of a ${}^{20}\mathrm{Ne}$ beam off a hydrogen target at an energy of 50~MeV/u.

\item[Method]   A beam of fully stripped ${}^{20}$Ne ions was injected into the ESR storage ring at an energy of 50 MeV/u. The beam interacted with an internal hydrogen gas-jet target. An ultra-high vacuum compatible detector setup was installed around the gas jet inside the ring to measure the recoiling particles generated by nuclear reactions.

\item[Results]   Multiple reaction channels were observed during the experiment. In particular, we present the results from studies on elastic and inelastic scattering, as well as the neutron transfer reaction ${}^{20}\mathrm{Ne}(p,d){}^{19}\mathrm{Ne}^*$. The experimental data were compared to calculations that took into account the most significant excited states, using a coupled-reaction channel approach. A very good agreement with the experimental data was achieved.

\item[Conclusions]  The present results are the first demonstration of the investigation transfer reactions using detectors directly installed  in the ring. This provides an important proof-of-principle for prospective studies with far-from-stability radioactive beams in the future.

\end{description}

\end{abstract}
\maketitle

\section{Introduction}
Direct nuclear reactions are a powerful tool to investigate various nuclear properties. For example, elastic scattering gives access to the nuclear size and radial matter distribution \cite{ALKHAZOV197889,ALKHAZOV2002269,DOBROVOLSKY20061,ILIEVA20128}. Inelastic scattering provides the possibility to study the deformation and collective excited modes of nuclei,  such as giant resonances \cite{harakeh2001giant, Zamora201616}. This enables the extraction of important nuclear bulk properties, like the dipole polarizability and the nuclear-matter incompressibility \cite{PhysRevLett.107.062502, GARG201855}. Similarly,  one- or a few-nucleon transfer reactions provide valuable insights into the structure and clustering phenomena in nuclei \cite{PhysRevLett.104.192501, RevModPhys.90.035004}. Direct reactions have important implications in nuclear astrophysics. In particular, the capture reactions of neutrons, protons, and alpha particles determine the production and isotopic abundances of the elements in the Universe. Crucial nuclear reactions in the stellar media have a significant influence on the evolution of astrophysical scenarios, such as core-collapse supernovae and neutron-star mergers~\cite{JANKA200738, neutronmergers}. \par

In the past, nuclear reaction experiments commonly employed forward kinematics, in which light ions served as the projectile, and the detected reaction products were mainly beam-like particles. With the availability of radioactive-ion beams and new detection methods, there has been great interest in investigating nuclear reactions in inverse kinematics in the last few years. Novel techniques like solid hydrogen/deuterium targets \cite{PhysRevLett.118.262502}, implanted tritium targets \cite{PhysRevLett.105.252501} or active targets \cite{BAZIN2020103790} have made a large variety of accurate measurements possible using radioactive beams in inverse kinematics. However, these experiments were constrained by  the energy and angular resolutions of the detected recoils, as well as the background produced by reactions on target and beam contaminants. The storage-ring technique is an alternative method to achieve high-resolution measurements
of direct reactions in inverse kinematics at very low momentum transfer.  Pioneering experiments using in-ring detection in  storage rings have demonstrated many advantages over other techniques.   For instance, low-momentum transfer experiments using this technique provided unique measurements of nuclear-matter densities and giant resonances. \cite{vonSchmid2023, Zamora201616,PhysRevC.96.034617}. \par

The stored-beam technique is the basis of the EXL (\underline{ex}otic nuclei studied with \underline{l}ight-ion-induced reactions in storage rings) project \cite{fair,1402-4896-2015-T166-014004, peter_exl}. Over the last years, the project has been operated  at the  \underline{e}xperimental heavy-ion \underline{s}torage \underline{r}ing (ESR) \cite{FRANZKE198718} at the  GSI facility. New projects like CARME \cite{carme} and NECTAR \cite{nectar}  are employing a concept similar to EXL for investigating nuclear reactions of astrophysical interest at storage rings. Currently, the technique is only applicable to studies involving nuclei with half-lives exceeding seconds, which represents the time needed for electron cooling. However, novel cooling methods, such as laser cooling, will significantly improve the beam cooling time and momentum spread \cite{STECK2020103811}. \par

In this work, we present the results of a nuclear reaction experiment using a storage ring with a ${}^{20}$Ne beam and a hydrogen gas target. This measurement was conducted simultaneously with the experiment E087 (P. J. Woods) \cite{E087}  reported in Ref.~\cite{Doherty_2015}. Nevertheless, electronics and detection systems were completely independent, yet the results of these experiments are complementary.  The detector setup employed in the present measurements was the same as described in previous publications \cite{vonSchmid2023, Zamora201616,PhysRevC.96.034617}. It is important to mention that the detectors used in this experiment were directly installed in the ring, which is different from the configuration employed in the experiment E087 \cite{E087, Doherty_2015}. In that experiment, a silicon detector was mounted in a ``pocket'' located downstream the gas-jet target. Therefore, this work shows the first demonstration of silicon detectors installed directly in the ring employed in a transfer-reaction experiment. The experimental procedure is explained below.

\section{Experiment}
The experiment was conducted at the heavy-ion storage ring ESR at the GSI facility \cite{FRANZKE198718}. The UNILAC-SIS18 system delivered a ${}^{20}$Ne  beam that was injected into the ESR at an energy of 50~MeV/u. Each beam injection stored about $10^8$ particles in the ring. Electron cooling efficiently reduced the momentum spread of the injected beam to less than $\Delta p/p \sim 10^{-5}$ \cite{Steck2004357}. The stored and cooled ${}^{20}$Ne beam then impinged on a hydrogen gas-jet target approximately  $10^6$ times per second. The target consisted of a pure hydrogen (H$_2$) flux, perpendicular to the beam direction, at supersonic speed and a thickness of about $10^{13}$ part./$\text{cm}^2$.  The diameter of the gas jet at the interaction zone with the stored beam was determined to be 7~mm by measuring (without electron cooling) the beam energy-loss  as a function of the orbit bump \cite{mirkotesis,mitesis}.  The relatively low target thickness was effectively compensated by the revolution frequency of the stored ion beam, resulting in a significant increase in luminosity.  A luminosity on the order of $10^{27}$~$\text{cm}^{-2}\text{s}^{-1}$ was achieved in this experiment. \par

\begin{figure*}[!ht]
\centering
\includegraphics[width=0.9\textwidth]{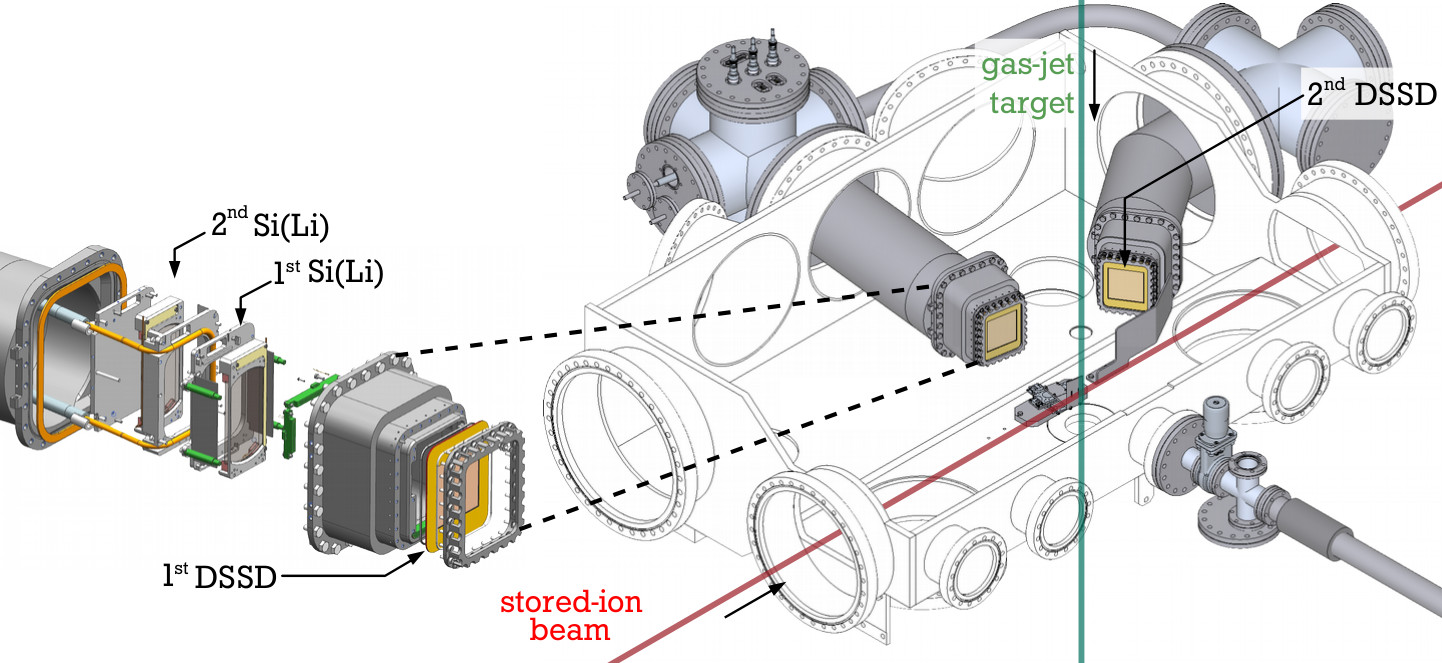}\\
\caption{\label{setup} Schematic illustration  of the vacuum chamber installed in the ESR for the present experiment \cite{mirkotesis}. The stored beam interacts with the vertical gas-jet target oriented perpendicular to the beam. The detectors were assembled at two internal pockets centered at 81$^\circ$ and 32\textdegree, with respect to the beam direction.  }
\end{figure*}

The detection system used in this experiment was especially designed for measurements at very low momentum transfer within a storage ring \cite{vonSchmid2023,Zamora201616,PhysRevC.96.034617}. The setup was ultra-high vacuum (UHV) [below $10^{-10}$~mbar] compatible, enabling measurements of light recoil particles (i.e., $p$, $d$, $t$, $^{3}$He and $\alpha$) with energies down to 100~keV. This was achieved by using double-sided silicon strip detectors (DSSDs) as {\it active vacuum windows} between the UHV environment of the ring and an auxiliary vacuum ($\sim 10^{-8}$~mbar) operating inside two internal {\it pockets} installed in the scattering chamber (see Fig.~\ref{setup}) \cite{Streicher2011604,1402-4896-2015-T166-014053}. The differential-vacuum concept allowed the use of subsequent layers of detectors inside the pockets, as well as all unbakeable,  outgassing electronic components and supporting materials. The DSSDs consist of $128\times64$ orthogonally-oriented strips, covering an area of $64\times64$~mm$^2$, with a total thickness of 285~$\mu$m. The position of the detectors was carefully selected to achieve measurements of elastic scattering and other reaction channels at very small center-of-mass angles. As seen in Fig.~\ref{setup}, a detector pocket was mounted at laboratory angles (relative to the beam direction) between 74$^\circ$ and 89$^\circ$, and the second pocket covered angles from 27$^\circ$ to 37$^\circ$. In addition, two thick (5~mm) cooled Si(Li) detectors were mounted behind the first DSSD (centered at 81$^\circ$) in order to stop elastically scattered protons that punch through the DSSD detector. \par
The angular resolution in this experiment was kinematically limited by the extension of the gas-jet target ($\sim 7$~mm diameter). The angular resolution achieved for this measurements was about 1$^\circ$ in the laboratory frame.  The use of a slit plate positioned in front of the detectors improved this value, resulting in a reduction of the angular resolution to 0.1$^\circ$ in the laboratory frame \cite{vonSchmid2023,PhysRevC.96.034617}. A Monte-Carlo simulation using the  \textsc{geant4} toolkit \cite{Agostinelli2003250} was performed to obtain the respective solid angle covered by each detector strip. Relative positions between the center of the target and  DSSDs were corrected by fitting  experimental spectra with \textsc{geant4}  simulations. Additionally,  a correlation between  DSSD pixels and  spherical coordinates  was derived from these simulations. This allowed to calibrate the  angular positions of each  strip (or pixel) with respect to  the  target center.

\section{Results \label{results}}

\begin{figure}[!ht]
\centering
\includegraphics[width=0.5\textwidth]{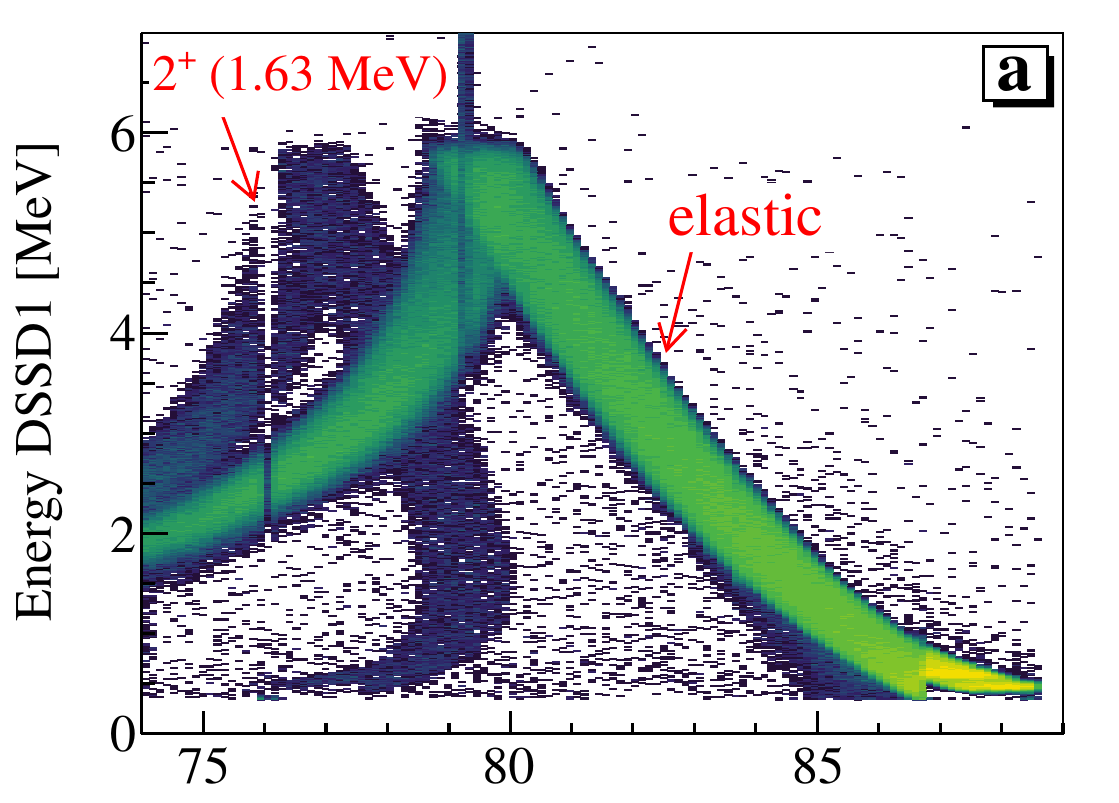}\\
\includegraphics[width=0.5\textwidth]{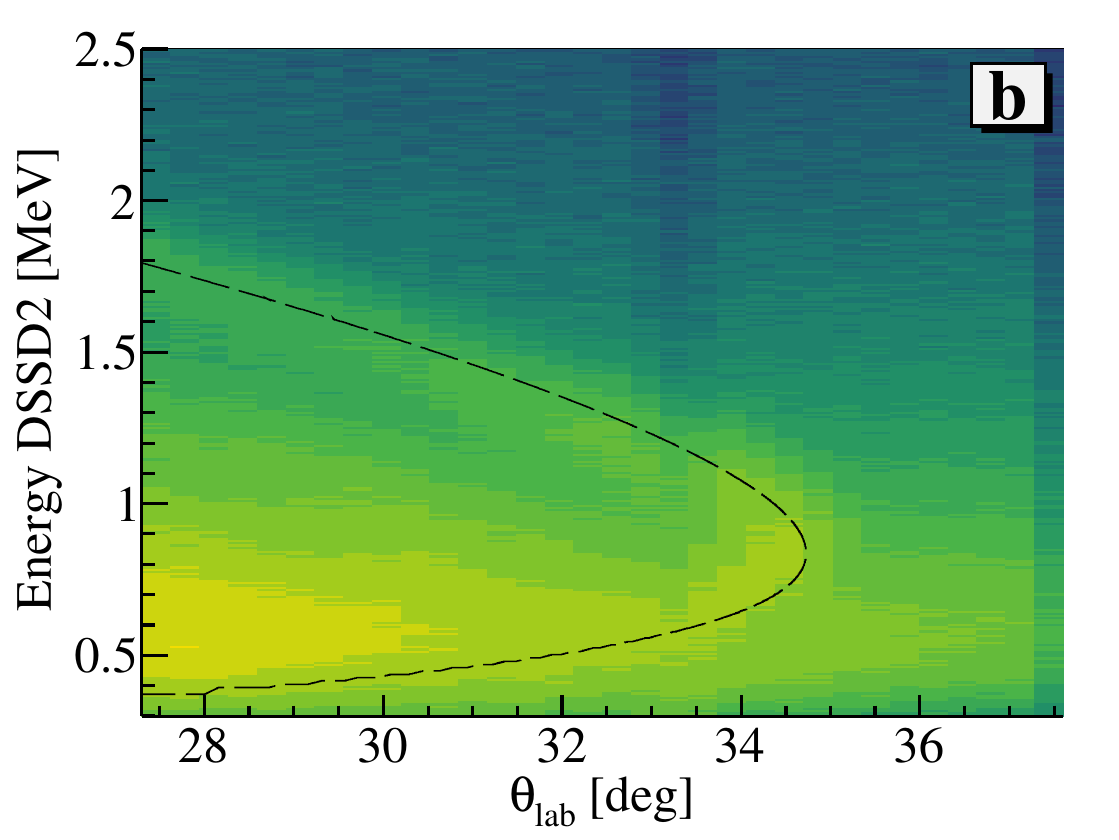}
\caption{\label{dssds}  Energy deposited by  recoils in the DSSD detectors vs.~laboratory polar angle corresponding to each strip. (a) Elastic and inelastic scattering measured with the detector DSSD1. To minimize systematic uncertainties in the reconstructed energy, the Si(Li) detectors were not used in the analysis. (b) One-neutron transfer reaction measured with the detector DSSD2. The kinematics is highlighted with a dashed line.}
\end{figure}

\begin{figure}[!ht]
\centering
\includegraphics[width=0.5\textwidth]{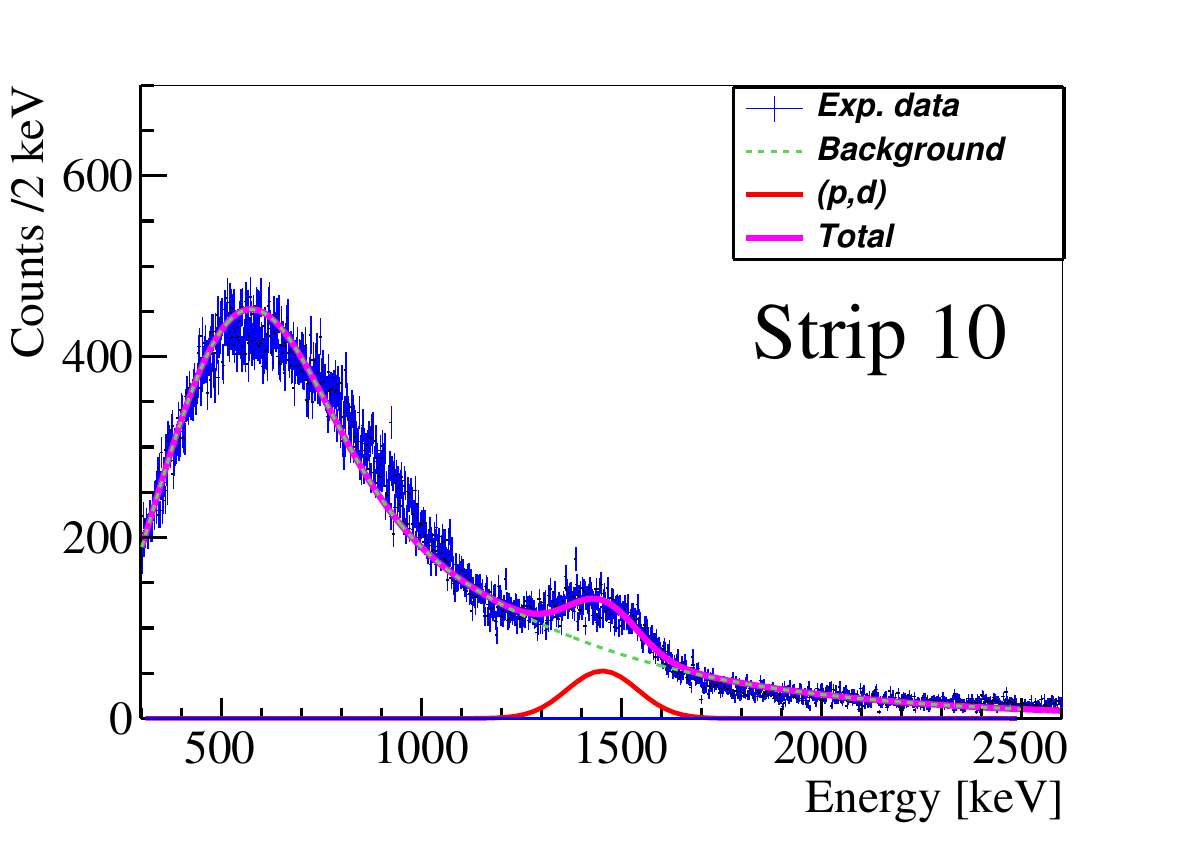}
\includegraphics[width=0.5\textwidth]{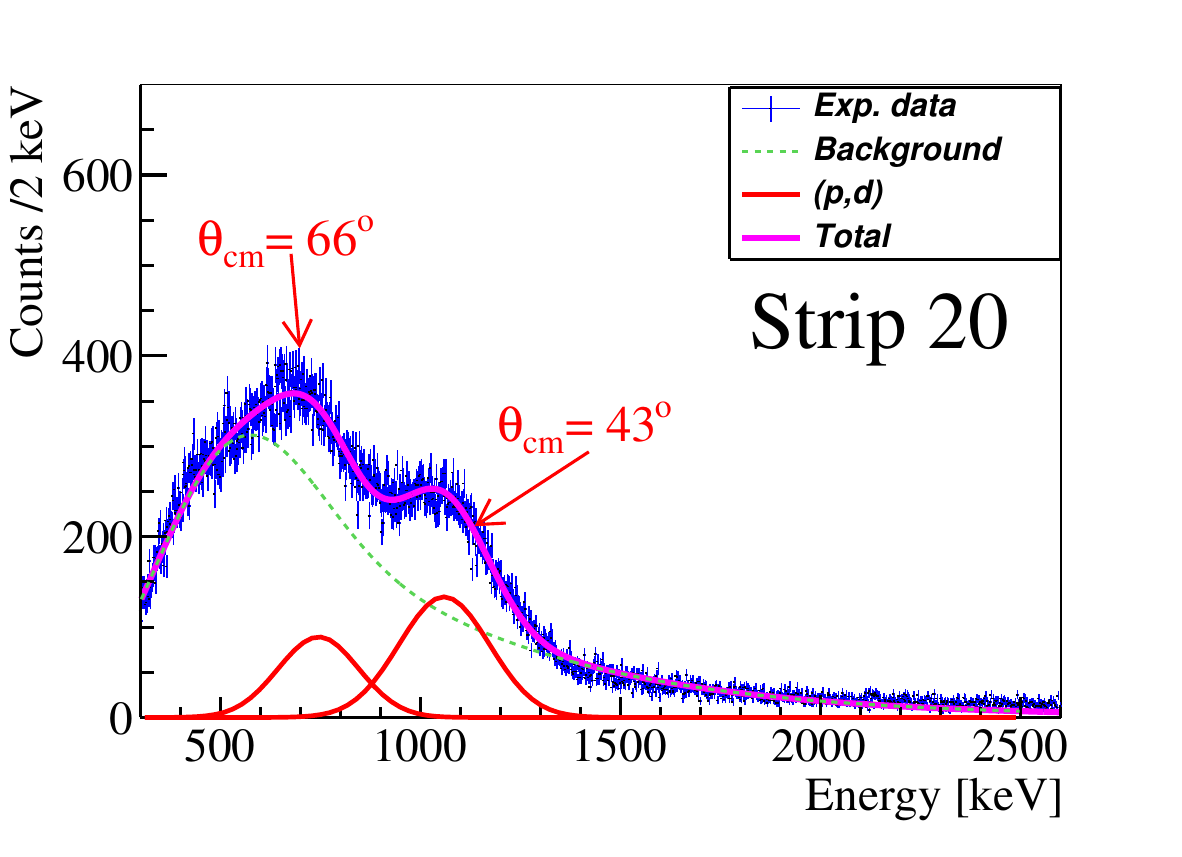}
\caption{\label{strip}  Energy spectra of strips number 10 and 20 ( around angles of 31$^\circ$ and 33$^\circ$ in the laboratory system) of the detector DSSD2. The experimental data clearly show bumps above the background, which correspond to transfer reaction channel. }
 \end{figure}

The present in-ring detection setup enabled  particle identification as well as an accurate determination of  the scattering angle and energy of light recoil particles.  Figure~\ref{dssds}(a) displays the measurement with the DSSD centered at 81$^\circ$ (labeled DSSD1) of elastic and inelastic scattering from the first $2^+$ excited state of ${}^{20}$Ne at 1.63~MeV. The kinematic plots exhibit an angular extension of about 1$^\circ$ as a result of the beam and target dimensions at the interaction point. The background is mainly produced from the scattering on residual gas particles in the chamber.
Simultaneously, a neutron transfer reaction ${}^{20}\mathrm{Ne}(p,d){}^{19}\mathrm{Ne}^*$ ($Q=-14.6$~MeV) was measured with the second DSSD (labeled DSSD2) located at the laboratory angle of 32$^\circ$. It should be noted that the detector position was not optimized for measuring this transfer reaction, as it was specifically placed for the measurement of an isoscalar giant resonance \cite{Zamora201616}. Due to the reaction kinematics, the scattering angle of the deuterons in the laboratory system is constrained to the forward region with a maximum value of 35$^\circ$. The kinetic energy of these deuterons in the angular coverage of the DSSD2 ranges from 30 to 130~MeV. Therefore, the detector thickness was not sufficient to fully stop the deuterons, and only a small fraction of their energy was deposited in the DSSD. Figure~\ref{dssds}(b) shows the ${}^{20}\mathrm{Ne}(p,d){}^{19}\mathrm{Ne}^*$ reaction measured in the DSSD2. The figure includes a dashed line representing the kinematic plot of the transfer reaction populating the ground state of ${}^{19}$Ne. In this case,  the primary source of background is predominantly derived from proton elastic and inelastic scattering, resulting in energy depositions below 1 MeV.
 Figure~\ref{strip} displays a projection of (vertical) strips 10 and 20, which correspond approximately to laboratory angles of 31$^\circ$ and 33$^\circ$, respectively. The background in all the strips was modeled using a log-normal function. The parameters of the function were adjusted based on the spectra of the last strips (around 37$^\circ$), which did not measure the transfer-reaction channel. The background shape remained constant, and the only parameter changed for each fit was the normalization. The transfer reaction is clearly observed since the centroid of the fitted peaks follows the kinematic line shown in Figure~\ref{dssds}(b). The energy resolution  was approximately 250~keV (full width at half maximum), as evident from the peaks observed at  center-of-mass angles of 43$^\circ$ and 66$^\circ$ in Fig.~\ref{strip}. Due to the kinematical compression of the reaction, it was not possible to disentangle the ground state and the energy levels below 2~MeV in ${}^{19}$Ne in this experiment. Hence, the measured peaks of the transfer reaction channel may have a contribution from a few low-lying states in ${}^{19}$Ne.  The average statistical error for each bin was 5\%, and it reached up to 15\% around the minima of the cross section.  The systematic error was approximately 16\%, with the uncertainty from the normalization being the main contributing component. \par

In order to investigate in detail the multiple reaction channels observed in this experiment, the coupled-reaction channels (CRC) formalism was employed for the analysis. The results are presented in the next Section.

\section{Theoretical Analysis \label{theo}}
\begin{figure}[htb!]
    \centering
    \includegraphics[width=0.45\textwidth]{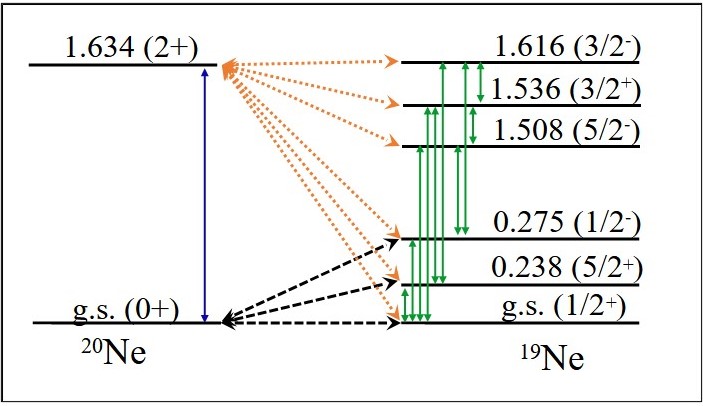}
    \caption{Coupling scheme considered for the projectile/ejectile overlap in the CRC calculation. Excited states were coupled assuming dipole and quadrupole electric (magnetic) transitions. }
    \label{coupled-scheme}
\end{figure}

Coupled-reaction channel calculations provide a  consistent method for studying the mechanisms  of diverse direct-reaction channels within the same model \cite{HAGINO2022103951}. This requires solving a coupled system of  Schr\"odinger equations that may involve projectile and target excitations and transfer reactions.  The computer code \textsc{fresco} \cite{THOMPSON1988167} has been used to perform the CRC calculations in this work. In the particular case of the ${}^{20}\mathrm{Ne}+p$ collision, elastic scattering and inelastic excitation are important reaction channels to include in the coupling scheme. Figure~\ref{coupled-scheme} shows the states of ${}^{20}$Ne and ${}^{19}$Ne that were considered for the CRC calculation. As can be seen, the ground and the first-excited state of ${}^{20}$Ne are used for the entrance partition to populate several low-lying states of ${}^{19}$Ne via  one-neutron transfer transitions (final partition). It is important to note that the coupling scheme also includes inelastic excitations in both projectile and ejectile nuclei. A detailed analysis of the CRC calculations and comparison to the experimental data is presented in the following Sections.

\subsection{Elastic and Inelastic Scattering}
The elastic scattering angular distribution was measured in center-of-mass angles ranging from 8$^\circ$ to 30$^\circ$, as shown in Fig.~\ref{elas-inelas}(a). The elastic scattering cross section is expressed as a ratio with the Rutherford cross section to eliminate the significant Coulomb dependence at forward angles. The theoretical description of this angular distribution is an essential step for the CRC calculation. In this work, the S\~{a}o Paulo (SP) potential  \cite{PhysRevLett.79.5218} was used to build the optical model (OM) potentials employed in the calculations. The SP potential is a parameter-free double-folding interaction widely used to describe the elastic scattering of nucleus-nucleus systems across a broad range of energies. Usually, the OM potential is built as
\begin{equation}
 U(r) = (N_R + i N_I)V_\mathrm{SP}(r),
\end{equation}
where $V_\mathrm{SP}(r)$ is the SP potential, $N_R$ and $N_I$ are the real and imaginary scaling factors, respectively. Systematical studies have shown that  elastic scattering angular distributions of a  large set of systems can be successfully reproduced by the scaling factors $N_R=1.0$ and $N_I=0.78$ \cite{078-2}. However,  the normalization of the imaginary part still deserves a more detailed analysis when the reaction involves light nuclei. In particular, for the ${}^{20}\mathrm{Ne}+p$ system, the OM potential that provides the best agreement with the experimental data was obtained by using a scaling factor of $N_I=0.5$ for the imaginary part. The difference between this value and the one derived from systematics can be attributed to the absence of reaction channels induced by a composite probe compared to protons. \\
The next step is incorporating couplings with other significant reaction channels, such as inelastic excitation and transfer reactions, into the calculation. As seen in the previous Section, the first  $2^+$ state  at 1.63 MeV of $^{20}$Ne was  strongly populated in the present experiment. This excited state has been shown to play an important role in the transfer reactions~\cite{FLL19,CFC20}. Following a similar approach, once inelastic and transfer channels were included in the coupled equations, the coefficient of the imaginary part  was reduced  to $N_I = 0.3$ in both partitions. This remanent potential accounts for other open channels that were not explicitly included in the coupling scheme of the reaction.  Notably, the imaginary coefficient utilized in the CRC calculation represents a reduction of approximately 40\% of the normalization employed to describe the elastic scattering channel. The CRC result for the elastic scattering is shown with the solid line in  Fig.~\ref{elas-inelas}(a). As can be seen, the calculation is in excellent agreement with the experimental data after including several couplings.  This procedure is consistent with the approach used in other works \cite{FCC21,PSV17,CFC20,FLL19, PhysRevC.106.014603}. The transition potential used for the analysis of the inelastic scattering angular distribution was based on the Bohr-Mottelson prescription \cite{bohr1969nuclear}:

\begin{equation}
 V_\lambda(r) =-\frac{\delta_\lambda}{4\pi}\frac{d}{dr}U(r),
\end{equation}
where $\lambda$ is the multipolatiry, $\delta_\lambda$ the deformation length and $U(r)$ is the OM potential employed for describing the elastic scattering data. Deformation of the Coulomb potential was also included in the analysis by using transition strengths derived from shell model calculations with the computer code \textsc{nushellx}  \cite{nushell}. Figure~\ref{elas-inelas}(b) displays the inelastic angular distribution measured for the $2_1^+$ state at 1.63~MeV in  $^{20}$Ne with the respective CRC result. In particular, $^{20}$Ne has a strong quadrupole deformation, which has been accurately determined through electron scattering experiments, resulting in a value of $\beta_2=0.73(3)$ \cite{RAMAN20011}. This value is in agreement with the value of $\beta_2=0.67$ used in the current analysis, which was obtained from \textsc{nushellx}.

\begin{figure}[htb!]
    \centering
    \includegraphics[width=0.5\textwidth]{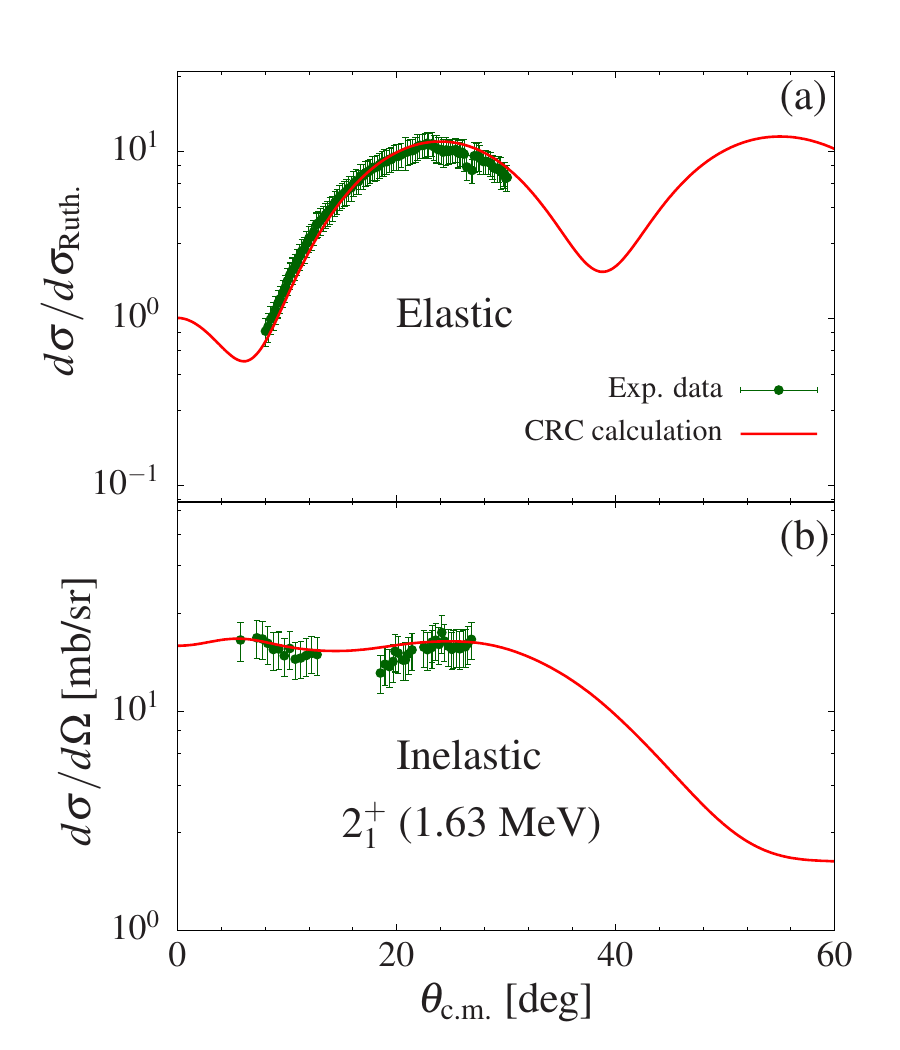}
    \caption{Angular distributions of elastic (a) and inelastic (b) scattering for the ${}^{20}\mathrm{Ne}+p$ system. Both statistical and systematic errors are included. The results of the CRC calculation are shown with a solid line. }
    \label{elas-inelas}
\end{figure}

\subsection{One Neutron Transfer}
\begin{figure}[htb!]
    \centering
    \includegraphics[width=0.5\textwidth]{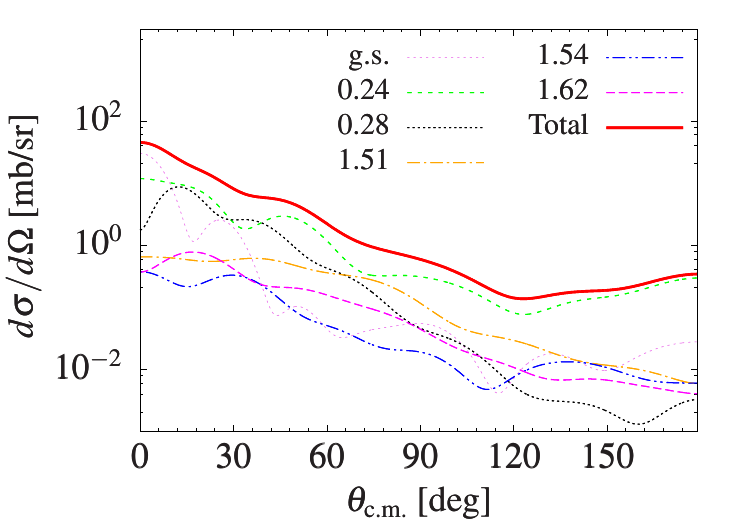}
    \caption{Differential cross sections obtained from the CRC calculation. Each line corresponds to an excited  state in ${}^{19}$Ne with the excitation energy in MeV. The solid line is the summed contribution of all states. }
    \label{TR_ad1}
\end{figure}

As indicated in Sect.~\ref{results}, the ${}^{20}\mathrm{Ne}(p,d){}^{19}\mathrm{Ne}^*$ transfer reaction was strongly populated in this experiment. Because of the kinematical compression due to the reaction mechanism in the laboratory system, it was not possible to disentangle states in ${}^{19}$Ne up to 2~MeV  in this experiment. Nevertheless, the CRC formalism offers a consistent approach for investigating the contribution of different states in the final partition. Dedicated shell-model calculations using the code \textsc{nushellx} were performed to obtain the respective spectroscopic amplitudes (SA) for the $\langle ^{20}\mathrm{Ne} \otimes \nu^{-1}_{njm}\vert ^{19}\mathrm{Ne}\rangle$ projectile/ejectile overlap. In this context, the shell-model Hamiltonian was constructed considering a model space containing the valence orbits $1p_{1/2}$, $1d_{5/2}$, and $2s_{1/2}$ for both protons and neutrons, for which the $^{12}$C nucleus is assumed as closed core. The single-particle energies were set to $\epsilon_{1p1/2} = -5.70$~MeV, $\epsilon_{1d5/2} = -1.67$~MeV, and $\epsilon_{2s1/2} = -2.84$~MeV, respectively.
The two-body matrix elements were determined by a least-squares fit procedure to reproduce the ground states and selected excited states of nuclei in the mass region $A = 13$--22 \cite{rewile}. The results obtained from shell-model calculations are presented in Table~\ref{amplitudes-20Ne-19Ne}. The spectroscopic amplitudes were used as input for the CRC calculation to achieve  a good description of the differential cross sections. Figure~\ref{TR_ad1} shows the angular distributions derived with the CRC calculation. As can be seen, the nuclear transitions populating the ${}^{19}$Ne ground state and excited states at 0.24 and 0.28~MeV are the main contributions to the total angular distribution. These results were verified through an additional analysis (see  Appendix~\label{app} A.) of the ${}^{20}\mathrm{Ne}(p,d){}^{19}\mathrm{Ne}^*$ reaction at 30~MeV, which was measured in forward kinematics and reported in Ref.~\cite{BCA17}. In particular, the transition to the $^{19}$Ne$_{\mathrm{0.24}}(5/2^{+})$ state becomes dominant for scattering angles $\theta_\text{c.m.} > 30^\circ$.

\begin{table}[h!]
\caption{Spectroscopic amplitudes for the projectile and ejectile overlap wave functions considered in the $p$($^{20}$Ne,$^{19}$Ne)$d$ reaction.}
\setlength{\tabcolsep}{6pt} 
\renewcommand{\arraystretch}{1.7} 
\centering
\begin{tabular}{cccc}
\hline\hline
Initial State  & $nl_{j}$ & Final State & S.A. \\ \hline
$^{20}$Ne$_{\mathrm{g.s.}}(0^{+})$ & $2s_{1/2}$ & $^{19}$Ne$_{\mathrm{g.s.}}(1/2^{+})$ &-0.81  \\
$^{20}$Ne$_{\mathrm{g.s.}}(0^{+})$ & $1d_{5/2}$ & $^{19}$Ne$_{\mathrm{0.24}}(5/2^{+})$ &-1.26  \\
$^{20}$Ne$_{\mathrm{g.s.}}(0^{+})$ & $1p_{1/2}$ & $^{19}$Ne$_{\mathrm{0.28}}(1/2^{-})$ &-1.19  \\
$^{20}$Ne$_{\mathrm{1.634}}(2^{+})$ & $1d_{5/2}$ & $^{19}$Ne$_{\mathrm{g.s.}}(1/2^{+})$ &-0.66  \\
$^{20}$Ne$_{\mathrm{1.634}}(2^{+})$ & $2s_{1/2}$ & $^{19}$Ne$_{\mathrm{0.24}}(5/2^{+})$ &-0.67  \\
$^{20}$Ne$_{\mathrm{1.634}}(2^{+})$ & $1d_{5/2}$ & $^{19}$Ne$_{\mathrm{0.24}}(5/2^{+})$ &-0.64  \\
$^{20}$Ne$_{\mathrm{1.634}}(2^{+})$ & $1p_{1/2}$ & $^{19}$Ne$_{\mathrm{1.51}}(5/2^{-})$ & 0.92  \\
$^{20}$Ne$_{\mathrm{1.634}}(2^{+})$ & $2s_{1/2}$ & $^{19}$Ne$_{\mathrm{1.54}}(3/2^{+})$ & 0.27  \\
$^{20}$Ne$_{\mathrm{1.634}}(2^{+})$ & $1d_{5/2}$ & $^{19}$Ne$_{\mathrm{1.54}}(3/2^{+})$ & 0.44  \\
$^{20}$Ne$_{\mathrm{1.634}}(2^{+})$ & $1p_{1/2}$ & $^{19}$Ne$_{\mathrm{1.62}}(3/2^{-})$ & 0.77  \\ \hline\hline
\end{tabular}
\label{amplitudes-20Ne-19Ne}
\end{table}

\begin{figure}[htb!]
    \centering
    \includegraphics[width=0.5\textwidth]{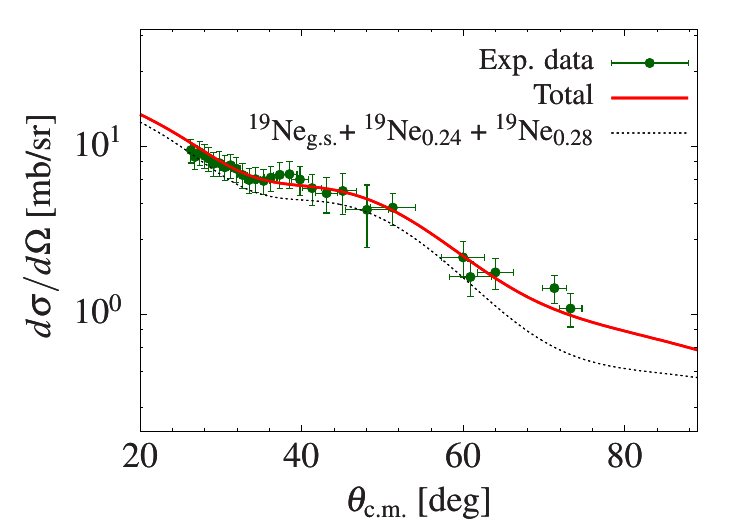}
    \caption{Transfer reaction channel measured with the detector DSSD2.  Both statistical and systematic errors are included. The solid line represents the result of the CRC calculation, illustrating the total contribution of the coupled states in ${}^{19}$Ne described in Fig.~\ref{coupled-scheme}. The dashed line represents only the summed contribution of the ground and the first two excited states in ${}^{19}$Ne.
    }
    \label{TR_ad2}
\end{figure}

A comparison of the CRC calculation with the experimental data measured in the present experiment is shown in Fig.~\ref{TR_ad2}. The total angular distribution, which includes six states in ${}^{19}$Ne below 2~MeV, is in very good agreement with the data. Similar to the results obtained from the analysis with the data from Ref.~\cite{BCA17}, the transfer reactions to the first three bound states of ${}^{19}$Ne are the main contribution to the measured cross section. However, the coupling with other possible excited states in the initial and final partitions is of great importance to fully describe all the angular distributions of this experiment.

\section{Summary}

 A nuclear reaction experiment was performed in a storage ring using a UHV-compatible detection system with a stored $^{20}$Ne beam and a hydrogen gas-jet target. The differential cross sections of the elastic and inelastic scattering reaction channels were obtained through measurements nearly free of background. Additionally, a transfer reaction was successfully measured for the first time using a detector system that was directly installed in the ring. The experimental data were compared to CRC calculations, assuming a consistent coupling scheme, including multiple excited states in both the projectile and ejectile nuclei. The theoretical results are in very good agreement with the present data.  This provides an important proof-of-principle for prospective studies  at small momentum transfer in storage rings with far-from-stability nuclei. New experiments are already planned for continuation with the physics program, including an extended detector setup covering more extensive angular ranges for nuclear reaction studies with unstable stored beams at GSI, and eventually at FAIR.

\section*{Acknowledgements}
\input acknowledgement.tex   

\appendix

\section{CRC calculation at 30 MeV \label{app}}

Calculations for an incoming energy of 30 MeV were performed to test the consistency of the CRC approach presented in this work. The same procedure described in Section~\label{theo} was used for the calculations. The CRC results were benchmarked with experimental data from literature. The elastic and inleastic scattering were extracted from Ref.~\cite{DESWINIARSKI197327}. The transfer reaction channel was benchmarked with data from Ref.~\cite{BCA17}. As can be seen, the experimental angular distributions for elastic and inelastic scattering are very well described by the CRC calculation. Also,  the calculated distributions for the ground state and the combination of the  0.24 and 0.28 MeV states are in good agreement with the data.

\begin{figure}[htb!]
    \centering
    \includegraphics[width=0.5\textwidth]{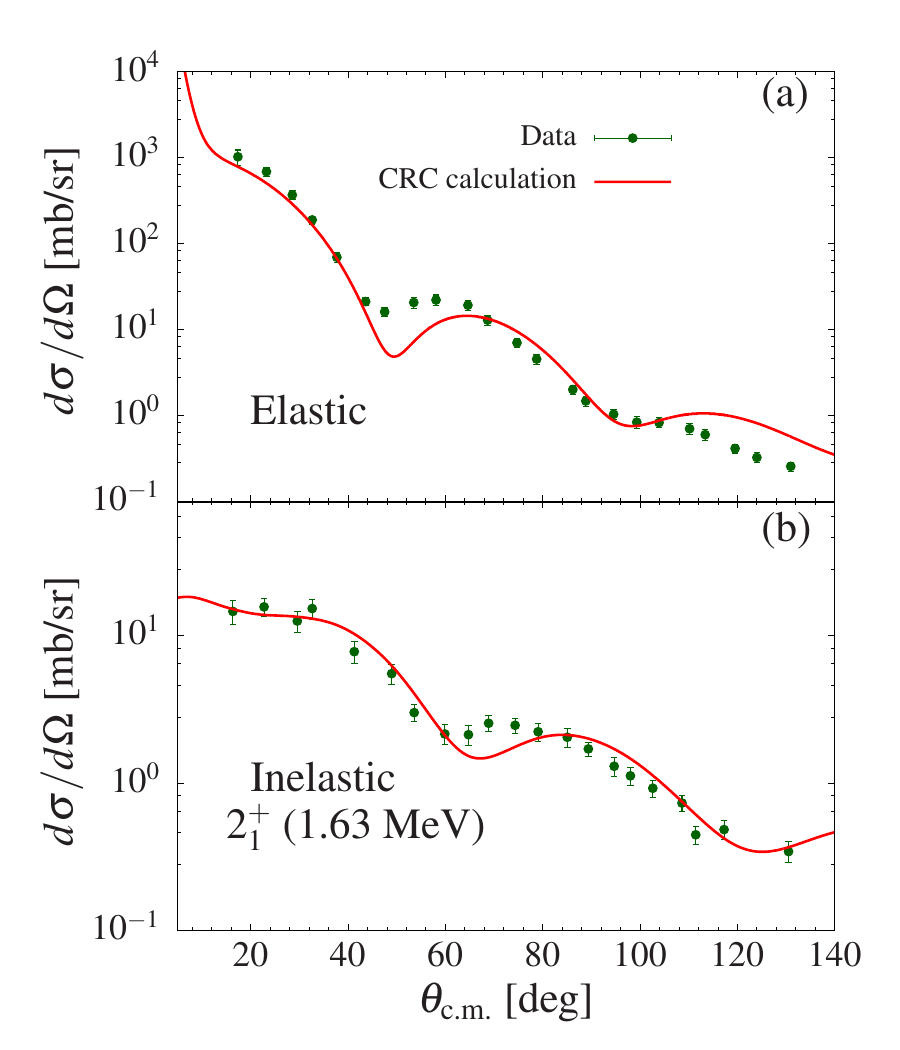}
    \caption{Angular distributions of elastic (a) and inelastic (b) scattering for the ${}^{20}\mathrm{Ne}+p$ system at 30 MeV.  The results of the CRC calculation are shown with a solid line. The data was extracted from Ref.~\cite{DESWINIARSKI197327}. }
    \label{elas-inelas}
\end{figure}

\begin{figure}[htb!]
    \centering
    \includegraphics[width=0.5\textwidth]{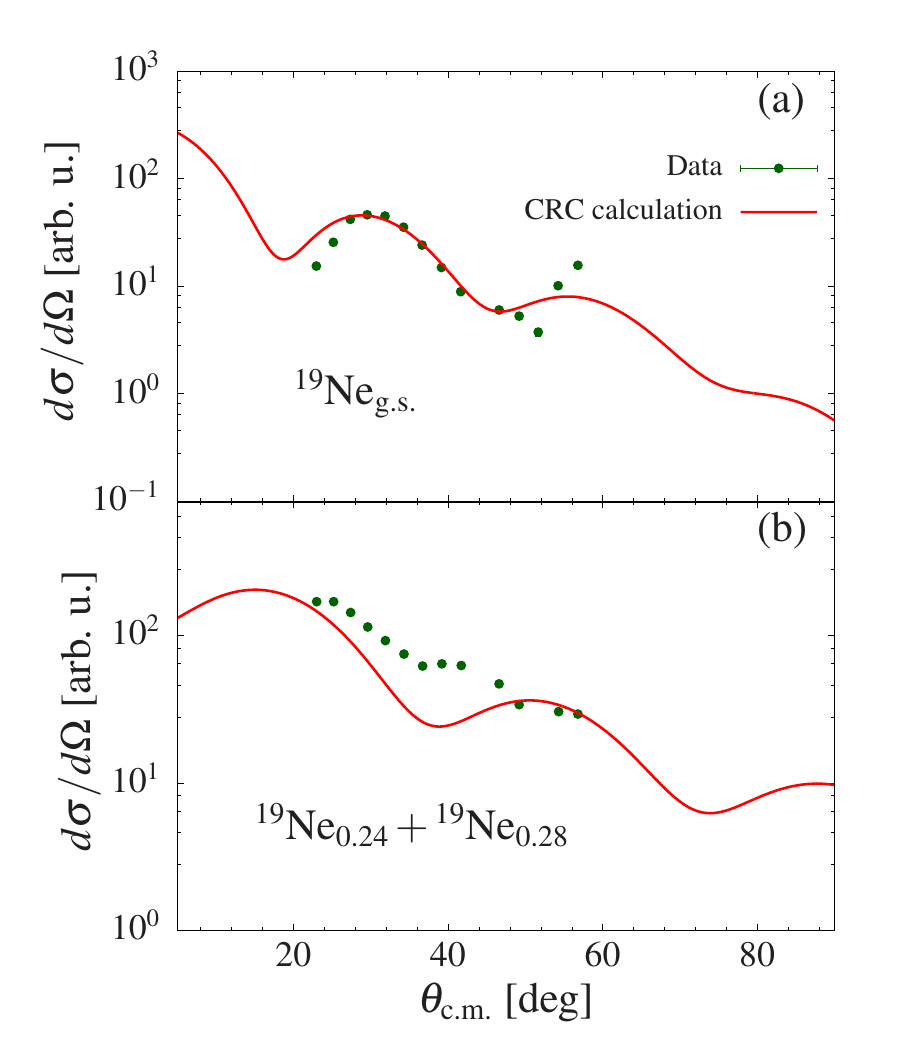}
    \caption{Angular distributions of the transfer reaction ${}^{20}\mathrm{Ne}(p,d){}^{19}\mathrm{Ne}^*$ at 30 MeV populating the ground state (a) and the sum of the 0.24 MeV ($L=2$) and 0.28 MeV ($L=1$) states in ${}^{19}$Ne. The results of the CRC calculation are shown with a solid line. The data was extracted from Ref.~\cite{BCA17}. The calculation results were scaled by a constant value to match the data. }
    \label{elas-inelas}
\end{figure}

\bibliography{bibliography}  

\end{document}

%% file: author_list.tex
\author{J. C.~Zamora}
\email[E-mail address: ]{zamora@frib.msu.edu}
\altaffiliation{Present address: Facility for Rare Isotope Beams (FRIB), Michigan State University, East Lansing, MI, USA}
\affiliation{Institut f\"{u}r Kernphysik, Technische Universit\"at Darmstadt, Germany}

\author{T.~Aumann}
\affiliation{Institut f\"{u}r Kernphysik, Technische Universit\"at Darmstadt, Germany}
\affiliation{GSI Helmholtzzentrum f\"ur Schwerionenforschung GmbH, Darmstadt, Germany}
\affiliation{Helmholtz Forschungsakademie Hessen f\"ur FAIR (HFHF), GSI Helmholtzzentrum für Schwerionenforschung, Campus Darmstadt, Germany}

\author{S.~Bagchi}
\altaffiliation{Present address: Department of Physics, IIT-ISM Dhanbad, Dhanbad, Jharkhand, India}
\affiliation{ESRIG, University of Groningen, The Netherlands}
\affiliation{GSI Helmholtzzentrum f\"ur Schwerionenforschung GmbH, Darmstadt, Germany}

\author{S.~Bishop}
\thanks{Deceased}
\affiliation{Physik-Department E12, Technische Universit\"at M\"unchen, Germany}

\author{M.~Bo}
\affiliation{Goethe-Universit\"at Frankfurt, Frankfurt, Germany}

\author{S.~B\"onig}
\affiliation{Institut f\"{u}r Kernphysik, Technische Universit\"at Darmstadt, Germany}

\author{C.~Brandau}
\affiliation{GSI Helmholtzzentrum f\"ur Schwerionenforschung GmbH, Darmstadt, Germany}
\affiliation{I. Physikalisches Institut, Justus-Liebig-Universit\"at Giessen, Germany}

\author{ M.~Csatl\'os}
\affiliation{Institute for Nuclear Research, HUN-REN ATOMKI, Debrecen, Hungary}

\author{T.~Davinson}
\affiliation{Institute for Particle and Nuclear Physics, University of Edinburgh, UK}

\author{I.~Dillmann}
\affiliation{GSI Helmholtzzentrum f\"ur Schwerionenforschung GmbH, Darmstadt, Germany}
\affiliation{II. Physikalisches Institut, Justus-Liebig-Universit\"at Giessen, Germany}

\author{C.~Dimopoulou}
\affiliation{GSI Helmholtzzentrum f\"ur Schwerionenforschung GmbH, Darmstadt, Germany}

\author{D.~T.~Doherty}
\affiliation{Institute for Particle and Nuclear Physics, University of Edinburgh, UK}

\author{P.~Egelhof}
\affiliation{GSI Helmholtzzentrum f\"ur Schwerionenforschung GmbH, Darmstadt, Germany}

\author{V.~Eremin}
\affiliation{Ioffe Physical-Technical Institute, St.~Petersburg, Russia}

\author{A.~Estrade}
\affiliation{Institute for Particle and Nuclear Physics, University of Edinburgh, UK}

\author{A.~Evdokimovc}
\affiliation{GSI Helmholtzzentrum f\"ur Schwerionenforschung GmbH, Darmstadt, Germany}
\affiliation{II. Physikalisches Institut, Justus-Liebig-Universit\"at Giessen, Germany}

\author{J. L.~Ferreira}
  \affiliation{Instituto de F\'isica, Universidade Federal Fluminense, Niter\'oi,  RJ, Brazil}

\author{T.~Furuno}
\affiliation{Division of Physics and Astronomy, Kyoto University, Japan}

\author{H.~Geissel}
\affiliation{GSI Helmholtzzentrum f\"ur Schwerionenforschung GmbH, Darmstadt, Germany}
\affiliation{II. Physikalisches Institut, Justus-Liebig-Universit\"at Giessen, Germany}

\author{R.~Gernh\"auser}
\affiliation{Physik-Department E12, Technische Universit\"at M\"unchen, Germany}

\author{A.~Gumberidze}
\affiliation{GSI Helmholtzzentrum f\"ur Schwerionenforschung GmbH, Darmstadt, Germany}

\author{M.~N.~Harakeh}
\affiliation{ESRIG, University of Groningen, The Netherlands}

\author{A.-L.~Hartig}
\affiliation{Institut f\"{u}r Kernphysik, Technische Universit\"at Darmstadt, Germany}

\author{M.~Heil}
\affiliation{GSI Helmholtzzentrum f\"ur Schwerionenforschung GmbH, Darmstadt, Germany}

\author{S.~Ilieva}
\affiliation{Institut f\"{u}r Kernphysik, Technische Universit\"at Darmstadt, Germany}

\author{N.~Kalantar-Nayestanaki}
\affiliation{ESRIG, University of Groningen, The Netherlands}

\author{O.~Kiselev}
\affiliation{GSI Helmholtzzentrum f\"ur Schwerionenforschung GmbH, Darmstadt, Germany}

\author{H.~Kollmus}
\affiliation{GSI Helmholtzzentrum f\"ur Schwerionenforschung GmbH, Darmstadt, Germany}

\author{C.~Kozhuharov}
\affiliation{GSI Helmholtzzentrum f\"ur Schwerionenforschung GmbH, Darmstadt, Germany}

\author{A.~Krasznahorkay}
\affiliation{Institute for Nuclear Research, HUN-REN ATOMKI, Debrecen, Hungary}

\author{Th.~Kr\"oll}
\affiliation{Institut f\"{u}r Kernphysik, Technische Universit\"at Darmstadt, Germany}
\affiliation{Helmholtz Forschungsakademie Hessen f\"ur FAIR (HFHF), GSI Helmholtzzentrum für Schwerionenforschung, Campus Darmstadt, Germany}

\author{M.~Kuilman}
\affiliation{ESRIG, University of Groningen, The Netherlands}

\author{C.~Lederer-Woods}
\affiliation{Goethe-Universit\"at Frankfurt, Frankfurt, Germany}

\author{S.~Litvinov}
\affiliation{GSI Helmholtzzentrum f\"ur Schwerionenforschung GmbH, Darmstadt, Germany}

\author{Yu.~A.~Litvinov}
\affiliation{GSI Helmholtzzentrum f\"ur Schwerionenforschung GmbH, Darmstadt, Germany}

\author{G.~Lotay}
\affiliation{Institute for Particle and Nuclear Physics, University of Edinburgh, UK}

\author{J.~Lubian}
 \affiliation{Instituto de F\'isica, Universidade Federal Fluminense, Niter\'oi, RJ, Brazil}

\author{M.~Mahjour-Shafiei}
\affiliation{Department of Physics, University of Tehran, Iran}
\affiliation{ESRIG, University of Groningen, The Netherlands}

\author{M.~Mutterer}
\thanks{Deceased}
\affiliation{GSI Helmholtzzentrum f\"ur Schwerionenforschung GmbH, Darmstadt, Germany}

\author{D.~Nagae}
\affiliation{Department of Physics, University of Tsukuba, Japan}

\author{M.~A.~Najafi}
\affiliation{ESRIG, University of Groningen, The Netherlands}

\author{C.~Nociforo}
\affiliation{GSI Helmholtzzentrum f\"ur Schwerionenforschung GmbH, Darmstadt, Germany}

\author{F.~Nolden}
\thanks{Deceased}
\affiliation{GSI Helmholtzzentrum f\"ur Schwerionenforschung GmbH, Darmstadt, Germany}

\author{N.~Petridis}
\affiliation{GSI Helmholtzzentrum f\"ur Schwerionenforschung GmbH, Darmstadt, Germany}

\author{U.~Popp}
\affiliation{GSI Helmholtzzentrum f\"ur Schwerionenforschung GmbH, Darmstadt, Germany}

\author{R.~Reifarth}
\affiliation{Goethe-Universit\"at Frankfurt, Frankfurt, Germany}

\author{C.~Rigollet}
\affiliation{ESRIG, University of Groningen, The Netherlands}

\author{S.~Roy}
\affiliation{ESRIG, University of Groningen, The Netherlands}

\author{C.~Scheidenberger}
\affiliation{GSI Helmholtzzentrum f\"ur Schwerionenforschung GmbH, Darmstadt, Germany}
\affiliation{II. Physikalisches Institut, Justus-Liebig-Universit\"at Giessen, Germany}

\author{M.~von~Schmid}
\affiliation{Institut f\"{u}r Kernphysik, Technische Universit\"at Darmstadt, Germany}

\author{M.~Steck}
\affiliation{GSI Helmholtzzentrum f\"ur Schwerionenforschung GmbH, Darmstadt, Germany}

\author{Th.~St\"ohlker}
\affiliation{GSI Helmholtzzentrum f\"ur Schwerionenforschung GmbH, Darmstadt, Germany}
\affiliation{Friedrich-Schiller-Universit\"at Jena, Jena, Germany}

\author{B.~Streicher}
\affiliation{GSI Helmholtzzentrum f\"ur Schwerionenforschung GmbH, Darmstadt, Germany}

\author{L.~Stuhl}
\affiliation{Institute for Nuclear Research, HUN-REN ATOMKI, Debrecen, Hungary}

\author{M.~Th\"urauf}
\affiliation{Institut f\"{u}r Kernphysik, Technische Universit\"at Darmstadt, Germany}

\author{S.~Trotsenko}
\affiliation{GSI Helmholtzzentrum f\"ur Schwerionenforschung GmbH, Darmstadt, Germany}

\author{T.~Uesaka}
\affiliation{RIKEN Nishina Center, Wako, Saitama, Japan}

\author{H.~Weick}
\affiliation{GSI Helmholtzzentrum f\"ur Schwerionenforschung GmbH, Darmstadt, Germany}

\author{J.~S.~Winfield}
\thanks{Deceased}
\affiliation{GSI Helmholtzzentrum f\"ur Schwerionenforschung GmbH, Darmstadt, Germany}

\author{D.~Winters}
\affiliation{GSI Helmholtzzentrum f\"ur Schwerionenforschung GmbH, Darmstadt, Germany}

\author{P.~J.~Woods}
\affiliation{Institute for Particle and Nuclear Physics, University of Edinburgh, UK}

\author{T.~Yamaguchi}
\affiliation{Department of Physics, Saitama University, Sakura-ku, Saitama, Japan}

\author{X.~L.~Yan}
\affiliation{GSI Helmholtzzentrum f\"ur Schwerionenforschung GmbH, Darmstadt, Germany}
\affiliation{Institute of Modern Physics, Chinese Academy of Sciences, Lanzhou, China}

\author{K.~Yue}
\affiliation{Institut f\"{u}r Kernphysik, Technische Universit\"at Darmstadt, Germany}
\affiliation{GSI Helmholtzzentrum f\"ur Schwerionenforschung GmbH, Darmstadt, Germany}
\affiliation{Institute of Modern Physics, Chinese Academy of Sciences, Lanzhou, China}

\author{J.~Zenihiro}
\affiliation{RIKEN Nishina Center, Wako, Saitama, Japan}


%% file: acknowledgement.tex
The results presented in this paper are based on work performed before Feb 24th 2022. We acknowledge technical support by A.~Glazenborg-Kluttig, M.~Lindemulder, P.~Schakel, H.~Timersma (ESRIG, Groningen), J.~Cavaco, G.~May,  L.~Urban  and the accelerator staff (GSI, Darmstadt).\\
This work was supported by German BMBF (06DA9040I, 05P12RDFN8, 05P15RDFN1, 05P19RGFA1 and 05P21RGFA1), the European Commission within the Seventh Framework Programme through IA-ENSAR (contract No. RII3-CT-2010-262010), the Hungarian OTKA Foundation No.\ K106035, the Sumitomo Foundation, the National Natural Science Foundation of China (contract No. 11575269), the HGF through the Helmholtz-CAS Joint Research Group HCJRG-108, HIC for FAIR, GSI-RUG/KVI collaboration agreement, TU Darmstadt-GSI cooperation contract and the STIBET Doctoral program of the DAAD.
Brazilian authors acknowledge partial financial support from CNPq, FAPERJ, CAPES, and INCT-FNA (Instituto Nacional de Ci\^{e}ncia e Tecnologia - F\'{i}sica Nuclear e Aplica\c{c}\~{o}es) research project 464898/2014-5.

%% file: article_20Ne_SR_rev.bbl
\begin{thebibliography}{47}%
\makeatletter
\providecommand \@ifxundefined [1]{%
 \@ifx{#1\undefined}
}%
\providecommand \@ifnum [1]{%
 \ifnum #1\expandafter \@firstoftwo
 \else \expandafter \@secondoftwo
 \fi
}%
\providecommand \@ifx [1]{%
 \ifx #1\expandafter \@firstoftwo
 \else \expandafter \@secondoftwo
 \fi
}%
\providecommand \natexlab [1]{#1}%
\providecommand \enquote  [1]{``#1''}%
\providecommand \bibnamefont  [1]{#1}%
\providecommand \bibfnamefont [1]{#1}%
\providecommand \citenamefont [1]{#1}%
\providecommand \href@noop [0]{\@secondoftwo}%
\providecommand \href [0]{\begingroup \@sanitize@url \@href}%
\providecommand \@href[1]{\@@startlink{#1}\@@href}%
\providecommand \@@href[1]{\endgroup#1\@@endlink}%
\providecommand \@sanitize@url [0]{\catcode `\\12\catcode `\$12\catcode
  `\&12\catcode `\#12\catcode `\^12\catcode `\_12\catcode `\%12\relax}%
\providecommand \@@startlink[1]{}%
\providecommand \@@endlink[0]{}%
\providecommand \url  [0]{\begingroup\@sanitize@url \@url }%
\providecommand \@url [1]{\endgroup\@href {#1}{\urlprefix }}%
\providecommand \urlprefix  [0]{URL }%
\providecommand \Eprint [0]{\href }%
\providecommand \doibase [0]{http://dx.doi.org/}%
\providecommand \selectlanguage [0]{\@gobble}%
\providecommand \bibinfo  [0]{\@secondoftwo}%
\providecommand \bibfield  [0]{\@secondoftwo}%
\providecommand \translation [1]{[#1]}%
\providecommand \BibitemOpen [0]{}%
\providecommand \bibitemStop [0]{}%
\providecommand \bibitemNoStop [0]{.\EOS\space}%
\providecommand \EOS [0]{\spacefactor3000\relax}%
\providecommand \BibitemShut  [1]{\csname bibitem#1\endcsname}%
\let\auto@bib@innerbib\@empty
\bibitem [{\citenamefont {Alkhazov}\ \emph {et~al.}(1978)\citenamefont
  {Alkhazov} \emph {et~al.}}]{ALKHAZOV197889}%
  \BibitemOpen
  \bibfield  {author} {\bibinfo {author} {\bibfnamefont {G.~D.}\ \bibnamefont
  {Alkhazov}} \emph {et~al.},\ }\href {\doibase
  http://dx.doi.org/10.1016/0370-1573(78)90083-2} {\bibfield  {journal}
  {\bibinfo  {journal} {Phys. Rep.}\ }\textbf {\bibinfo {volume} {42}},\
  \bibinfo {pages} {89 } (\bibinfo {year} {1978})}\BibitemShut {NoStop}%
\bibitem [{\citenamefont {Alkhazov}\ \emph {et~al.}(2002)\citenamefont
  {Alkhazov} \emph {et~al.}}]{ALKHAZOV2002269}%
  \BibitemOpen
  \bibfield  {author} {\bibinfo {author} {\bibfnamefont {G.}~\bibnamefont
  {Alkhazov}} \emph {et~al.},\ }\href {\doibase
  http://dx.doi.org/10.1016/S0375-9474(02)01273-3} {\bibfield  {journal}
  {\bibinfo  {journal} {Nucl. Phys. A}\ }\textbf {\bibinfo {volume} {712}},\
  \bibinfo {pages} {269 } (\bibinfo {year} {2002})}\BibitemShut {NoStop}%
\bibitem [{\citenamefont {Dobrovolsky}\ \emph {et~al.}(2006)\citenamefont
  {Dobrovolsky} \emph {et~al.}}]{DOBROVOLSKY20061}%
  \BibitemOpen
  \bibfield  {author} {\bibinfo {author} {\bibfnamefont {A.}~\bibnamefont
  {Dobrovolsky}} \emph {et~al.},\ }\href {\doibase
  http://dx.doi.org/10.1016/j.nuclphysa.2005.11.016} {\bibfield  {journal}
  {\bibinfo  {journal} {Nucl. Phys. A}\ }\textbf {\bibinfo {volume} {766}},\
  \bibinfo {pages} {1 } (\bibinfo {year} {2006})}\BibitemShut {NoStop}%
\bibitem [{\citenamefont {Ilieva}\ \emph {et~al.}(2012)\citenamefont {Ilieva}
  \emph {et~al.}}]{ILIEVA20128}%
  \BibitemOpen
  \bibfield  {author} {\bibinfo {author} {\bibfnamefont {S.}~\bibnamefont
  {Ilieva}} \emph {et~al.},\ }\href {\doibase
  http://dx.doi.org/10.1016/j.nuclphysa.2011.11.010} {\bibfield  {journal}
  {\bibinfo  {journal} {Nucl. Phys. A}\ }\textbf {\bibinfo {volume} {875}},\
  \bibinfo {pages} {8 } (\bibinfo {year} {2012})}\BibitemShut {NoStop}%
\bibitem [{\citenamefont {Harakeh}\ and\ \citenamefont {van~der
  Woude}(2001)}]{harakeh2001giant}%
  \BibitemOpen
  \bibfield  {author} {\bibinfo {author} {\bibfnamefont {M.~N.}\ \bibnamefont
  {Harakeh}}\ and\ \bibinfo {author} {\bibfnamefont {A.}~\bibnamefont {van~der
  Woude}},\ }\href@noop {} {\emph {\bibinfo {title} {Giant Resonances:
  Fundamental High-frequency Modes of Nuclear Excitation}}},\ Oxford science
  publications\ (\bibinfo  {publisher} {Oxford University Press},\ \bibinfo
  {year} {2001})\BibitemShut {NoStop}%
\bibitem [{\citenamefont {Zamora}\ \emph {et~al.}(2016)\citenamefont {Zamora}
  \emph {et~al.}}]{Zamora201616}%
  \BibitemOpen
  \bibfield  {author} {\bibinfo {author} {\bibfnamefont {J.~C.}\ \bibnamefont
  {Zamora}} \emph {et~al.},\ }\href {\doibase
  http://dx.doi.org/10.1016/j.physletb.2016.10.015} {\bibfield  {journal}
  {\bibinfo  {journal} {Phys. Lett. B}\ }\textbf {\bibinfo {volume} {763}},\
  \bibinfo {pages} {16 } (\bibinfo {year} {2016})}\BibitemShut {NoStop}%
\bibitem [{\citenamefont {Tamii}\ \emph {et~al.}(2011)\citenamefont {Tamii}
  \emph {et~al.}}]{PhysRevLett.107.062502}%
  \BibitemOpen
  \bibfield  {author} {\bibinfo {author} {\bibfnamefont {A.}~\bibnamefont
  {Tamii}} \emph {et~al.},\ }\href {\doibase 10.1103/PhysRevLett.107.062502}
  {\bibfield  {journal} {\bibinfo  {journal} {Phys. Rev. Lett.}\ }\textbf
  {\bibinfo {volume} {107}},\ \bibinfo {pages} {062502} (\bibinfo {year}
  {2011})}\BibitemShut {NoStop}%
\bibitem [{\citenamefont {Garg}\ and\ \citenamefont
  {Col\`o}(2018)}]{GARG201855}%
  \BibitemOpen
  \bibfield  {author} {\bibinfo {author} {\bibfnamefont {U.}~\bibnamefont
  {Garg}}\ and\ \bibinfo {author} {\bibfnamefont {G.}~\bibnamefont {Col\`o}},\
  }\href {\doibase https://doi.org/10.1016/j.ppnp.2018.03.001} {\bibfield
  {journal} {\bibinfo  {journal} {Prog. Part. Nucl. Phys}\ }\textbf {\bibinfo
  {volume} {101}},\ \bibinfo {pages} {55} (\bibinfo {year} {2018})}\BibitemShut
  {NoStop}%
\bibitem [{\citenamefont {Catford}\ \emph {et~al.}(2010)\citenamefont {Catford}
  \emph {et~al.}}]{PhysRevLett.104.192501}%
  \BibitemOpen
  \bibfield  {author} {\bibinfo {author} {\bibfnamefont {W.~N.}\ \bibnamefont
  {Catford}} \emph {et~al.},\ }\href {\doibase 10.1103/PhysRevLett.104.192501}
  {\bibfield  {journal} {\bibinfo  {journal} {Phys. Rev. Lett.}\ }\textbf
  {\bibinfo {volume} {104}},\ \bibinfo {pages} {192501} (\bibinfo {year}
  {2010})}\BibitemShut {NoStop}%
\bibitem [{\citenamefont {Freer}\ \emph {et~al.}(2018)\citenamefont {Freer},
  \citenamefont {Horiuchi}, \citenamefont {Kanada-En'yo}, \citenamefont {Lee},\
  and\ \citenamefont {Mei\ss{}ner}}]{RevModPhys.90.035004}%
  \BibitemOpen
  \bibfield  {author} {\bibinfo {author} {\bibfnamefont {M.}~\bibnamefont
  {Freer}}, \bibinfo {author} {\bibfnamefont {H.}~\bibnamefont {Horiuchi}},
  \bibinfo {author} {\bibfnamefont {Y.}~\bibnamefont {Kanada-En'yo}}, \bibinfo
  {author} {\bibfnamefont {D.}~\bibnamefont {Lee}}, \ and\ \bibinfo {author}
  {\bibfnamefont {U.-G.}\ \bibnamefont {Mei\ss{}ner}},\ }\href {\doibase
  10.1103/RevModPhys.90.035004} {\bibfield  {journal} {\bibinfo  {journal}
  {Rev. Mod. Phys.}\ }\textbf {\bibinfo {volume} {90}},\ \bibinfo {pages}
  {035004} (\bibinfo {year} {2018})}\BibitemShut {NoStop}%
\bibitem [{\citenamefont {Janka}\ \emph {et~al.}(2007)\citenamefont {Janka},
  \citenamefont {Langanke}, \citenamefont {Marek}, \citenamefont
  {Martínez-Pinedo},\ and\ \citenamefont {M\:uller}}]{JANKA200738}%
  \BibitemOpen
  \bibfield  {author} {\bibinfo {author} {\bibfnamefont {H.-T.}\ \bibnamefont
  {Janka}}, \bibinfo {author} {\bibfnamefont {K.}~\bibnamefont {Langanke}},
  \bibinfo {author} {\bibfnamefont {A.}~\bibnamefont {Marek}}, \bibinfo
  {author} {\bibfnamefont {G.}~\bibnamefont {Martínez-Pinedo}}, \ and\
  \bibinfo {author} {\bibfnamefont {B.}~\bibnamefont {M\:uller}},\ }\href
  {\doibase https://doi.org/10.1016/j.physrep.2007.02.002} {\bibfield
  {journal} {\bibinfo  {journal} {Phys. Rep.}\ }\textbf {\bibinfo {volume}
  {442}},\ \bibinfo {pages} {38} (\bibinfo {year} {2007})},\ \bibinfo {note}
  {the Hans Bethe Centennial Volume 1906-2006}\BibitemShut {NoStop}%
\bibitem [{\citenamefont {Thielemann}\ \emph {et~al.}(2017)\citenamefont
  {Thielemann}, \citenamefont {Eichler}, \citenamefont {Panov},\ and\
  \citenamefont {Wehmeyer}}]{neutronmergers}%
  \BibitemOpen
  \bibfield  {author} {\bibinfo {author} {\bibfnamefont {F.-K.}\ \bibnamefont
  {Thielemann}}, \bibinfo {author} {\bibfnamefont {M.}~\bibnamefont {Eichler}},
  \bibinfo {author} {\bibfnamefont {I.}~\bibnamefont {Panov}}, \ and\ \bibinfo
  {author} {\bibfnamefont {B.}~\bibnamefont {Wehmeyer}},\ }\href {\doibase
  10.1146/annurev-nucl-101916-123246} {\bibfield  {journal} {\bibinfo
  {journal} {Annu. Rev. Nucl.}\ }\textbf {\bibinfo {volume} {67}},\ \bibinfo
  {pages} {253} (\bibinfo {year} {2017})},\ \Eprint
  {http://arxiv.org/abs/https://doi.org/10.1146/annurev-nucl-101916-123246}
  {https://doi.org/10.1146/annurev-nucl-101916-123246} \BibitemShut {NoStop}%
\bibitem [{\citenamefont {Kumar}\ \emph {et~al.}(2017)\citenamefont {Kumar}
  \emph {et~al.}}]{PhysRevLett.118.262502}%
  \BibitemOpen
  \bibfield  {author} {\bibinfo {author} {\bibfnamefont {A.}~\bibnamefont
  {Kumar}} \emph {et~al.},\ }\href {\doibase 10.1103/PhysRevLett.118.262502}
  {\bibfield  {journal} {\bibinfo  {journal} {Phys. Rev. Lett.}\ }\textbf
  {\bibinfo {volume} {118}},\ \bibinfo {pages} {262502} (\bibinfo {year}
  {2017})}\BibitemShut {NoStop}%
\bibitem [{\citenamefont {Wimmer}\ \emph {et~al.}(2010)\citenamefont {Wimmer}
  \emph {et~al.}}]{PhysRevLett.105.252501}%
  \BibitemOpen
  \bibfield  {author} {\bibinfo {author} {\bibfnamefont {K.}~\bibnamefont
  {Wimmer}} \emph {et~al.},\ }\href {\doibase 10.1103/PhysRevLett.105.252501}
  {\bibfield  {journal} {\bibinfo  {journal} {Phys. Rev. Lett.}\ }\textbf
  {\bibinfo {volume} {105}},\ \bibinfo {pages} {252501} (\bibinfo {year}
  {2010})}\BibitemShut {NoStop}%
\bibitem [{\citenamefont {Bazin}\ \emph {et~al.}(2020)\citenamefont {Bazin}
  \emph {et~al.}}]{BAZIN2020103790}%
  \BibitemOpen
  \bibfield  {author} {\bibinfo {author} {\bibfnamefont {D.}~\bibnamefont
  {Bazin}} \emph {et~al.},\ }\href {\doibase
  https://doi.org/10.1016/j.ppnp.2020.103790} {\bibfield  {journal} {\bibinfo
  {journal} {Prog. Part. Nucl. Phys.}\ }\textbf {\bibinfo {volume} {114}},\
  \bibinfo {pages} {103790} (\bibinfo {year} {2020})}\BibitemShut {NoStop}%
\bibitem [{\citenamefont {von Schmid}\ \emph {et~al.}(2023)\citenamefont {von
  Schmid} \emph {et~al.}}]{vonSchmid2023}%
  \BibitemOpen
  \bibfield  {author} {\bibinfo {author} {\bibfnamefont {M.}~\bibnamefont {von
  Schmid}} \emph {et~al.},\ }\href {\doibase 10.1140/epja/s10050-023-00967-z}
  {\bibfield  {journal} {\bibinfo  {journal} {Eur. Phys. J. A}\ }\textbf
  {\bibinfo {volume} {59}},\ \bibinfo {pages} {83} (\bibinfo {year}
  {2023})}\BibitemShut {NoStop}%
\bibitem [{\citenamefont {Zamora}\ \emph {et~al.}(2017)\citenamefont {Zamora}
  \emph {et~al.}}]{PhysRevC.96.034617}%
  \BibitemOpen
  \bibfield  {author} {\bibinfo {author} {\bibfnamefont {J.~C.}\ \bibnamefont
  {Zamora}} \emph {et~al.},\ }\href {\doibase 10.1103/PhysRevC.96.034617}
  {\bibfield  {journal} {\bibinfo  {journal} {Phys. Rev. C}\ }\textbf {\bibinfo
  {volume} {96}},\ \bibinfo {pages} {034617} (\bibinfo {year}
  {2017})}\BibitemShut {NoStop}%
\bibitem [{\citenamefont {Gutbrod}\ \emph {et~al.}(2006)\citenamefont {Gutbrod}
  \emph {et~al.}}]{fair}%
  \BibitemOpen
  \bibfield  {author} {\bibinfo {author} {\bibfnamefont {H.~H.}\ \bibnamefont
  {Gutbrod}} \emph {et~al.},\ }\href@noop {} {\emph {\bibinfo {title} {FAIR
  Baseline Technical Report}}},\ ISBN 3-9811298-0-6\ (\bibinfo {year}
  {2006})\BibitemShut {NoStop}%
\bibitem [{\citenamefont {Kiselev}(2015)}]{1402-4896-2015-T166-014004}%
  \BibitemOpen
  \bibfield  {author} {\bibinfo {author} {\bibfnamefont {O.~A.}\ \bibnamefont
  {Kiselev}},\ }\href {http://stacks.iop.org/1402-4896/2015/i=T166/a=014004}
  {\bibfield  {journal} {\bibinfo  {journal} {Phys. Scr.}\ }\textbf {\bibinfo
  {volume} {T166}},\ \bibinfo {pages} {014004} (\bibinfo {year}
  {2015})}\BibitemShut {NoStop}%
\bibitem [{\citenamefont {Egelhof}()}]{peter_exl}%
  \BibitemOpen
  \bibfield  {author} {\bibinfo {author} {\bibfnamefont {P.}~\bibnamefont
  {Egelhof}},\ }\bibfield  {booktitle} {\emph {\bibinfo {booktitle} {Proc.
  STORI17}},\ }\href {\doibase 10.7566/JPSCP.35.011002} {\
  10.7566/JPSCP.35.011002},\ \Eprint
  {http://arxiv.org/abs/https://journals.jps.jp/doi/pdf/10.7566/JPSCP.35.011002}
  {https://journals.jps.jp/doi/pdf/10.7566/JPSCP.35.011002} \BibitemShut
  {NoStop}%
\bibitem [{\citenamefont {Franzke}(1987)}]{FRANZKE198718}%
  \BibitemOpen
  \bibfield  {author} {\bibinfo {author} {\bibfnamefont {B.}~\bibnamefont
  {Franzke}},\ }\href {\doibase http://dx.doi.org/10.1016/0168-583X(87)90583-0}
  {\bibfield  {journal} {\bibinfo  {journal} {Nucl. Instrum. Meth. Phys. Res.
  B}\ }\textbf {\bibinfo {volume} {24}},\ \bibinfo {pages} {18 } (\bibinfo
  {year} {1987})}\BibitemShut {NoStop}%
\bibitem [{\citenamefont {Bruno}\ \emph {et~al.}(2023)\citenamefont {Bruno}
  \emph {et~al.}}]{carme}%
  \BibitemOpen
  \bibfield  {author} {\bibinfo {author} {\bibfnamefont {C.}~\bibnamefont
  {Bruno}} \emph {et~al.},\ }\href {\doibase
  https://doi.org/10.1016/j.nima.2022.168007} {\bibfield  {journal} {\bibinfo
  {journal} {Nucl. Instrum. Meth. Phys. Res. A}\ }\textbf {\bibinfo {volume}
  {1048}},\ \bibinfo {pages} {168007} (\bibinfo {year} {2023})}\BibitemShut
  {NoStop}%
\bibitem [{\citenamefont {Jurado}(2023)}]{nectar}%
  \BibitemOpen
  \bibfield  {author} {\bibinfo {author} {\bibfnamefont {B.}~\bibnamefont
  {Jurado}},\ }\href {\doibase 10.1080/10619127.2023.2230849} {\bibfield
  {journal} {\bibinfo  {journal} {Nucl. Phys. News}\ }\textbf {\bibinfo
  {volume} {33}},\ \bibinfo {pages} {19} (\bibinfo {year} {2023})},\ \Eprint
  {http://arxiv.org/abs/https://doi.org/10.1080/10619127.2023.2230849}
  {https://doi.org/10.1080/10619127.2023.2230849} \BibitemShut {NoStop}%
\bibitem [{\citenamefont {Steck}\ and\ \citenamefont
  {Litvinov}(2020)}]{STECK2020103811}%
  \BibitemOpen
  \bibfield  {author} {\bibinfo {author} {\bibfnamefont {M.}~\bibnamefont
  {Steck}}\ and\ \bibinfo {author} {\bibfnamefont {Y.~A.}\ \bibnamefont
  {Litvinov}},\ }\href {\doibase https://doi.org/10.1016/j.ppnp.2020.103811}
  {\bibfield  {journal} {\bibinfo  {journal} {Prog. Part. Nucl. Phys.}\
  }\textbf {\bibinfo {volume} {115}},\ \bibinfo {pages} {103811} (\bibinfo
  {year} {2020})}\BibitemShut {NoStop}%
\bibitem [{E08()}]{E087}%
  \BibitemOpen
  \href@noop {} {\enquote {\bibinfo {title} {{Breakout of the Hot CNO Cycle in
  X-ray Bursts.}}}\ }\bibinfo {note} {{GSI Proposal (E087), 2010. P. J. Woods
  and Y. A. Litvinov}}\BibitemShut {NoStop}%
\bibitem [{\citenamefont {Doherty}\ \emph {et~al.}(2015)\citenamefont {Doherty}
  \emph {et~al.}}]{Doherty_2015}%
  \BibitemOpen
  \bibfield  {author} {\bibinfo {author} {\bibfnamefont {D.~T.}\ \bibnamefont
  {Doherty}} \emph {et~al.},\ }\href {\doibase
  10.1088/0031-8949/2015/T166/014007} {\bibfield  {journal} {\bibinfo
  {journal} {Phys. Scr.}\ }\textbf {\bibinfo {volume} {2015}},\ \bibinfo
  {pages} {014007} (\bibinfo {year} {2015})}\BibitemShut {NoStop}%
\bibitem [{\citenamefont {Steck}\ \emph {et~al.}(2004)\citenamefont {Steck}
  \emph {et~al.}}]{Steck2004357}%
  \BibitemOpen
  \bibfield  {author} {\bibinfo {author} {\bibfnamefont {M.}~\bibnamefont
  {Steck}} \emph {et~al.},\ }\href@noop {} {\bibfield  {journal} {\bibinfo
  {journal} {Nucl. Instrum. Meth. Phys. Res. A}\ }\textbf {\bibinfo {volume}
  {532}},\ \bibinfo {pages} {357} (\bibinfo {year} {2004})}\BibitemShut
  {NoStop}%
\bibitem [{\citenamefont {von Schmid}(2015)}]{mirkotesis}%
  \BibitemOpen
  \bibfield  {author} {\bibinfo {author} {\bibfnamefont {M.}~\bibnamefont {von
  Schmid}},\ }\emph {\bibinfo {title} {Nuclear matter distribution of 56Ni
  measured with EXL}},\ \href@noop {} {Ph.D. thesis},\ \bibinfo  {school} {TU
  Darmstadt} (\bibinfo {year} {2015})\BibitemShut {NoStop}%
\bibitem [{\citenamefont {Zamora}(2015)}]{mitesis}%
  \BibitemOpen
  \bibfield  {author} {\bibinfo {author} {\bibfnamefont {J.~C.}\ \bibnamefont
  {Zamora}},\ }\emph {\bibinfo {title} {Nuclear Reaction Studies using Stored
  Ion Beams at ESR with EXL}},\ \href@noop {} {Ph.D. thesis},\ \bibinfo
  {school} {TU Darmstadt} (\bibinfo {year} {2015})\BibitemShut {NoStop}%
\bibitem [{\citenamefont {Streicher}\ \emph {et~al.}(2011)\citenamefont
  {Streicher} \emph {et~al.}}]{Streicher2011604}%
  \BibitemOpen
  \bibfield  {author} {\bibinfo {author} {\bibfnamefont {B.}~\bibnamefont
  {Streicher}} \emph {et~al.},\ }\href@noop {} {\bibfield  {journal} {\bibinfo
  {journal} {Nucl. Instrum. Meth. Phys. Res. A}\ }\textbf {\bibinfo {volume}
  {654}},\ \bibinfo {pages} {604} (\bibinfo {year} {2011})}\BibitemShut
  {NoStop}%
\bibitem [{\citenamefont {Mutterer}\ \emph {et~al.}(2015)\citenamefont
  {Mutterer} \emph {et~al.}}]{1402-4896-2015-T166-014053}%
  \BibitemOpen
  \bibfield  {author} {\bibinfo {author} {\bibfnamefont {M.}~\bibnamefont
  {Mutterer}} \emph {et~al.},\ }\href
  {http://stacks.iop.org/1402-4896/2015/i=T166/a=014053} {\bibfield  {journal}
  {\bibinfo  {journal} {Phys. Scr.}\ }\textbf {\bibinfo {volume} {T166}},\
  \bibinfo {pages} {014053} (\bibinfo {year} {2015})}\BibitemShut {NoStop}%
\bibitem [{\citenamefont {Agostinelli}\ \emph {et~al.}(2003)\citenamefont
  {Agostinelli} \emph {et~al.}}]{Agostinelli2003250}%
  \BibitemOpen
  \bibfield  {author} {\bibinfo {author} {\bibfnamefont {S.}~\bibnamefont
  {Agostinelli}} \emph {et~al.},\ }\href {\doibase
  http://dx.doi.org/10.1016/S0168-9002(03)01368-8} {\bibfield  {journal}
  {\bibinfo  {journal} {Nucl. Instrum. Meth. Phys. Res. A}\ }\textbf {\bibinfo
  {volume} {506}},\ \bibinfo {pages} {250 } (\bibinfo {year}
  {2003})}\BibitemShut {NoStop}%
\bibitem [{\citenamefont {Hagino}\ \emph {et~al.}(2022)\citenamefont {Hagino},
  \citenamefont {Ogata},\ and\ \citenamefont {Moro}}]{HAGINO2022103951}%
  \BibitemOpen
  \bibfield  {author} {\bibinfo {author} {\bibfnamefont {K.}~\bibnamefont
  {Hagino}}, \bibinfo {author} {\bibfnamefont {K.}~\bibnamefont {Ogata}}, \
  and\ \bibinfo {author} {\bibfnamefont {A.}~\bibnamefont {Moro}},\ }\href
  {\doibase https://doi.org/10.1016/j.ppnp.2022.103951} {\bibfield  {journal}
  {\bibinfo  {journal} {Prog. Part. Nucl. Phys.}\ }\textbf {\bibinfo {volume}
  {125}},\ \bibinfo {pages} {103951} (\bibinfo {year} {2022})}\BibitemShut
  {NoStop}%
\bibitem [{\citenamefont {Thompson}(1988)}]{THOMPSON1988167}%
  \BibitemOpen
  \bibfield  {author} {\bibinfo {author} {\bibfnamefont {I.~J.}\ \bibnamefont
  {Thompson}},\ }\href {\doibase https://doi.org/10.1016/0167-7977(88)90005-6}
  {\bibfield  {journal} {\bibinfo  {journal} {Comput. Phys. Rep.}\ }\textbf
  {\bibinfo {volume} {7}},\ \bibinfo {pages} {167} (\bibinfo {year}
  {1988})}\BibitemShut {NoStop}%
\bibitem [{\citenamefont {Chamon}\ \emph {et~al.}(1997)\citenamefont {Chamon}
  \emph {et~al.}}]{PhysRevLett.79.5218}%
  \BibitemOpen
  \bibfield  {author} {\bibinfo {author} {\bibfnamefont {L.~C.}\ \bibnamefont
  {Chamon}} \emph {et~al.},\ }\href {\doibase 10.1103/PhysRevLett.79.5218}
  {\bibfield  {journal} {\bibinfo  {journal} {Phys. Rev. Lett.}\ }\textbf
  {\bibinfo {volume} {79}},\ \bibinfo {pages} {5218} (\bibinfo {year}
  {1997})}\BibitemShut {NoStop}%
\bibitem [{\citenamefont {Gasques}\ \emph {et~al.}(2006)\citenamefont
  {Gasques}, \citenamefont {Chamon}, \citenamefont {Gomes},\ and\ \citenamefont
  {Lubian}}]{078-2}%
  \BibitemOpen
  \bibfield  {author} {\bibinfo {author} {\bibfnamefont {L.}~\bibnamefont
  {Gasques}}, \bibinfo {author} {\bibfnamefont {L.}~\bibnamefont {Chamon}},
  \bibinfo {author} {\bibfnamefont {P.}~\bibnamefont {Gomes}}, \ and\ \bibinfo
  {author} {\bibfnamefont {J.}~\bibnamefont {Lubian}},\ }\href {\doibase
  https://doi.org/10.1016/j.nuclphysa.2005.09.001} {\bibfield  {journal}
  {\bibinfo  {journal} {Nucl. Phys. A}\ }\textbf {\bibinfo {volume} {764}},\
  \bibinfo {pages} {135} (\bibinfo {year} {2006})}\BibitemShut {NoStop}%
\bibitem [{\citenamefont {Ferreira}\ \emph {et~al.}(2019)\citenamefont
  {Ferreira} \emph {et~al.}}]{FLL19}%
  \BibitemOpen
  \bibfield  {author} {\bibinfo {author} {\bibfnamefont {J.~L.}\ \bibnamefont
  {Ferreira}} \emph {et~al.},\ }\href {\doibase 10.1140/epja/i2019-12773-7}
  {\bibfield  {journal} {\bibinfo  {journal} {Eur. Phys. J. A}\ }\textbf
  {\bibinfo {volume} {55}},\ \bibinfo {pages} {94} (\bibinfo {year}
  {2019})}\BibitemShut {NoStop}%
\bibitem [{\citenamefont {Carbone}\ \emph {et~al.}(2020)\citenamefont {Carbone}
  \emph {et~al.}}]{CFC20}%
  \BibitemOpen
  \bibfield  {author} {\bibinfo {author} {\bibfnamefont {D.}~\bibnamefont
  {Carbone}} \emph {et~al.},\ }\href@noop {} {\bibfield  {journal} {\bibinfo
  {journal} {Phys. Rev C}\ }\textbf {\bibinfo {volume} {102}} (\bibinfo {year}
  {2020})}\BibitemShut {NoStop}%
\bibitem [{\citenamefont {Ferreira}\ \emph {et~al.}(2021)\citenamefont
  {Ferreira} \emph {et~al.}}]{FCC21}%
  \BibitemOpen
  \bibfield  {author} {\bibinfo {author} {\bibfnamefont {J.~L.}\ \bibnamefont
  {Ferreira}} \emph {et~al.} (\bibinfo {collaboration} {NUMEN Collaboration}),\
  }\href {\doibase 10.1103/PhysRevC.103.054604} {\bibfield  {journal} {\bibinfo
   {journal} {Phys. Rev. C}\ }\textbf {\bibinfo {volume} {103}},\ \bibinfo
  {pages} {054604} (\bibinfo {year} {2021})}\BibitemShut {NoStop}%
\bibitem [{\citenamefont {Paes}\ \emph {et~al.}(2017)\citenamefont {Paes} \emph
  {et~al.}}]{PSV17}%
  \BibitemOpen
  \bibfield  {author} {\bibinfo {author} {\bibfnamefont {B.}~\bibnamefont
  {Paes}} \emph {et~al.},\ }\href@noop {} {\bibfield  {journal} {\bibinfo
  {journal} {Phys. Rev. C}\ }\textbf {\bibinfo {volume} {96}},\ \bibinfo
  {pages} {044612} (\bibinfo {year} {2017})}\BibitemShut {NoStop}%
\bibitem [{\citenamefont {Zamora}\ \emph {et~al.}(2022)\citenamefont {Zamora}
  \emph {et~al.}}]{PhysRevC.106.014603}%
  \BibitemOpen
  \bibfield  {author} {\bibinfo {author} {\bibfnamefont {J.~C.}\ \bibnamefont
  {Zamora}} \emph {et~al.},\ }\href {\doibase 10.1103/PhysRevC.106.014603}
  {\bibfield  {journal} {\bibinfo  {journal} {Phys. Rev. C}\ }\textbf {\bibinfo
  {volume} {106}},\ \bibinfo {pages} {014603} (\bibinfo {year}
  {2022})}\BibitemShut {NoStop}%
\bibitem [{\citenamefont {Bohr}\ and\ \citenamefont
  {Mottelson}(1975)}]{bohr1969nuclear}%
  \BibitemOpen
  \bibfield  {author} {\bibinfo {author} {\bibfnamefont {A.}~\bibnamefont
  {Bohr}}\ and\ \bibinfo {author} {\bibfnamefont {B.~R.}\ \bibnamefont
  {Mottelson}},\ }\href
  {http://inis.iaea.org/search/search.aspx?orig_q=RN:07247547} {\emph {\bibinfo
  {title} {Nuclear structure Volume II Nuclear deformations}}}\ (\bibinfo
  {publisher} {Addison-Wesley/W A Benjamin, Inc},\ \bibinfo {address} {United
  States},\ \bibinfo {year} {1975})\BibitemShut {NoStop}%
\bibitem [{nus()}]{nushell}%
  \BibitemOpen
  \href@noop {} {\enquote {\bibinfo {title} {{NuShellX for Windows and
  Linux}},}\ }\bibinfo {howpublished}
  {\url{http://www.garsington.eclipse.co.uk/}},\ \bibinfo {note} {accessed:
  2022-02-14}\BibitemShut {NoStop}%
\bibitem [{\citenamefont {Raman}\ \emph {et~al.}(2001)\citenamefont {Raman},
  \citenamefont {Nestor},\ and\ \citenamefont {Tikkanen}}]{RAMAN20011}%
  \BibitemOpen
  \bibfield  {author} {\bibinfo {author} {\bibfnamefont {S.}~\bibnamefont
  {Raman}}, \bibinfo {author} {\bibfnamefont {C.}~\bibnamefont {Nestor}}, \
  and\ \bibinfo {author} {\bibfnamefont {P.}~\bibnamefont {Tikkanen}},\ }\href
  {\doibase https://doi.org/10.1006/adnd.2001.0858} {\bibfield  {journal}
  {\bibinfo  {journal} {Atomic Data and Nuclear Data Tables}\ }\textbf
  {\bibinfo {volume} {78}},\ \bibinfo {pages} {1} (\bibinfo {year}
  {2001})}\BibitemShut {NoStop}%
\bibitem [{\citenamefont {McGrory}\ and\ \citenamefont
  {Wildenthal}(1973)}]{rewile}%
  \BibitemOpen
  \bibfield  {author} {\bibinfo {author} {\bibfnamefont {J.~B.}\ \bibnamefont
  {McGrory}}\ and\ \bibinfo {author} {\bibfnamefont {B.~H.}\ \bibnamefont
  {Wildenthal}},\ }\href {\doibase 10.1103/PhysRevC.7.974} {\bibfield
  {journal} {\bibinfo  {journal} {Phys. Rev. C}\ }\textbf {\bibinfo {volume}
  {7}},\ \bibinfo {pages} {974} (\bibinfo {year} {1973})}\BibitemShut {NoStop}%
\bibitem [{\citenamefont {Bardayan}\ \emph {et~al.}(2017)\citenamefont
  {Bardayan} \emph {et~al.}}]{BCA17}%
  \BibitemOpen
  \bibfield  {author} {\bibinfo {author} {\bibfnamefont {D.~W.}\ \bibnamefont
  {Bardayan}} \emph {et~al.},\ }\href@noop {} {\bibfield  {journal} {\bibinfo
  {journal} {Phys. Rev. C}\ }\textbf {\bibinfo {volume} {96}},\ \bibinfo
  {pages} {055806} (\bibinfo {year} {2017})}\BibitemShut {NoStop}%
\bibitem [{\citenamefont {{De Swiniarski}}\ \emph {et~al.}(1973)\citenamefont
  {{De Swiniarski}} \emph {et~al.}}]{DESWINIARSKI197327}%
  \BibitemOpen
  \bibfield  {author} {\bibinfo {author} {\bibfnamefont {R.}~\bibnamefont {{De
  Swiniarski}}} \emph {et~al.},\ }\href {\doibase
  https://doi.org/10.1016/0370-2693(73)90535-2} {\bibfield  {journal} {\bibinfo
   {journal} {Phys. Lett, B}\ }\textbf {\bibinfo {volume} {43}},\ \bibinfo
  {pages} {27} (\bibinfo {year} {1973})}\BibitemShut {NoStop}%
\end{thebibliography}%
